\documentclass[aps,prl,twocolumn,letterpaper]{revtex4-1} 

\usepackage[T1]{fontenc}
\usepackage[english,UKenglish]{babel}
\usepackage[utf8]{inputenc}
\usepackage{latexsym} 			
\usepackage{graphicx}			
\usepackage{color}    			
\usepackage[centertags]{amsmath} 	
\usepackage{amsxtra} 			
\usepackage{amsfonts} 			
\usepackage{amssymb}  			
\usepackage{bbm}      			
\usepackage{ae}				
\usepackage{type1cm}			
\usepackage{mathrsfs}			
\usepackage{enumitem}
\usepackage{hyphenat}			

\usepackage{MnSymbol}

\usepackage[bookmarks,			
            bookmarksopen=false,	
            bookmarksnumbered=true,	
            pdfstartview=Fit, 		
	    linkbordercolor={0 0 1},		
            citebordercolor={0 0.80 0.2},	
	    urlbordercolor={0.55 0.11 0.38},	
            colorlinks,			
            linkcolor=blue,	
	    citecolor=blue,	
            urlcolor=blue,	
	    filecolor=blue 		
            pdfauthor={Alexander Hentschel},
	    plainpages=false,
	    pdfpagelabels]{hyperref}	

\newcommand{\hs}[1] {\hspace{#1}}
\newcommand{\vs}[1] {\vspace{#1}}

\newcommand{\tn}[1]{\textnormal{#1}}

\newcommand{\bra}[1]{\langle #1 |}
\newcommand{\ket}[1]{| #1 \rangle}

\newcommand{\minfrac}[2]{\genfrac{}{}{}{1}{#1}{#2}}

\newcommand{\pEst}{\widetilde{\varphi}}

\newcommand{\swarmsize}{\Xi}

\newcommand{\policyPRL}{\bm{\varrho}}

\definecolor{Hhellgrau}{rgb}{0.8,0.8,0.8}
\definecolor{grey}{rgb}{0.6,0.6,0.6}
\definecolor{maroon}{rgb}{0.40,0.08,0.30}
\definecolor{darkred}{rgb}{0.80,0.0,0.0}
\definecolor{darkblue}{rgb}{0.0,0.0,0.55}
\definecolor{forestgreen}{rgb}{0.08,0.33,0.08}
\definecolor{darkgreen}{rgb}{0.00,0.40,0.00}
\definecolor{maroon}{rgb}{0.40,0.08,0.30}
\definecolor{brown}{rgb}{0.54, 0.32, 0.18}

\newcommand{\SupplementMaterial}[1]{}

\DeclareMathAlphabet{\mathitbf}{OML}{cmm}{b}{it}
\renewcommand{\bm}[1]{\begingroup\ensuremath{\boldsymbol #1}\endgroup}

\begin{document}


\addtolength{\topmargin}{0.125in}
\addtolength{\oddsidemargin}{0.00in}


\title{An Efficient Algorithm for Optimizing Adaptive Quantum Metrology Processes}
\author{Alexander Hentschel}
\author{Barry C.\ Sanders}
\affiliation{Institute for Quantum Information Science, University of Calgary, Calgary, Alberta, Canada T2N 1N4}

\begin{abstract}
\noindent
Quantum-enhanced metrology infers an unknown quantity with accuracy beyond the standard quantum limit (SQL). Feedback-based metrological techniques are promising for beating the SQL but devising the feedback procedures is difficult and inefficient. Here we introduce an efficient self-learning swarm-intelligence algorithm for devising feedback-based quantum metrological procedures. Our algorithm can be trained with simulated or real-world trials and accommodates experimental imperfections, losses, and decoherence.
\end{abstract}

\maketitle


\noindent
Precise metrology underpins modern science and engineering.
However, the `standard quantum limit' (SQL) restricts achievable precision, beyond which measurement must be treated on a quantum level.
Quantum-enhanced metrology (QEM) aims to beat the SQL by exploiting
entangled or squeezed input states and a sophisticated detection strategy 
\cite{giovannetti:010401,Pezze&Smerzi:PhysRevLett.2009,Luo:Lett.Math.Phys:2000}.
Feedback-based QEM is most effective as accumulated measurement data are exploited to maximize information gain in subsequent measurements, but finding an optimal QEM policy for a given measurement device is computationally intractable even for pure input states, unitary evolution $U$, and projective measurements.
Typically, policies have been devised by clever guessing \cite{PhysRevLett.85.5098,PhysRevA.63.053804} or brute-force numerical optimization \cite{PhysRevA.63.053804}.
Recently we introduced swarm-intelligence reinforcement learning
to devise optimal policies for measuring an interferometric phase shift \cite{QLearning:hentschel:PRL:2010}.
Our algorithm is space efficient;
i.e.\ the memory requirement is a polynomial function of the number of times $N$ that $U$ is 
effected,
in contrast to the exponentially expensive brute-force algorithm.
Although our result demonstrated the power of reinforcement learning,
our algorithm requires a run-time that is exponential in $N$ and a perfect interferometer, 
thereby effectively restricting its applicability to proofs of principle.
Here we report a space- \emph{and} time\hyp{}efficient algorithm 
(based on new heuristics) for devising QEM policies. 
Our algorithm works for noisy evolution and loss,
thus making reinforcement learning viable for autonomous
design of feedback-based QEM in a real-world setting.

We restrict our focus to single-parameter QEM. 
Interferometric phase estimation is the canonical quantum metrology problem
and is applicable to measurements of time, displacements, and imaging.
Therefore, we develop and benchmark our algorithm for autonomous policy design in this context.
To beat the SQL, we employ an entangled sequence of $N$ input photons, feedback control, and direct measurements of
the interferometer output.
For adaptive phase estimation, the interferometer processes one photon at a time. 
Each input photon can be in two modes, labeled $\{\ket{0},\ket{1}\}$, corresponding to the interferometer's two paths. 
Thus, a time-ordered sequence of $N$ photons implements an $N$-qubit state.

\begin{figure}
       \includegraphics[width=0.3\textwidth]{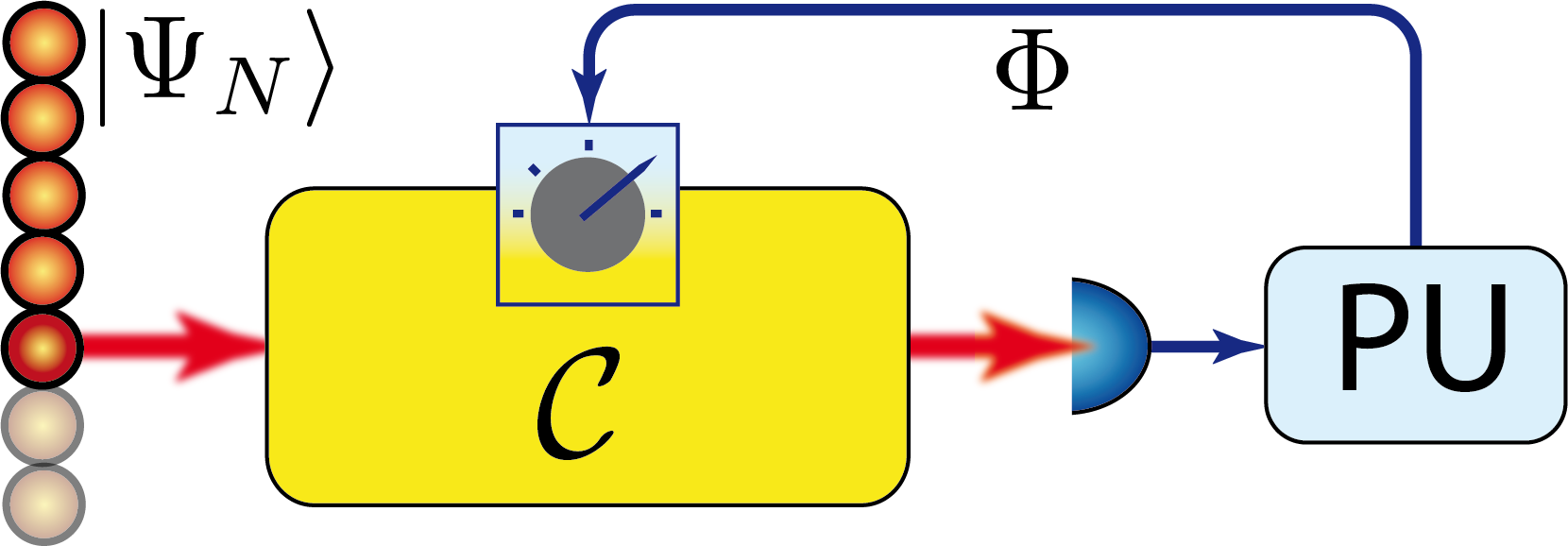}
	\caption[]{
		Adaptive feedback scheme for estimating an interferometric phase $\varphi$.
		The input state $\ket{\Psi_N}$ is fed into the unital quantum channel $\mathcal{C}$ one qubit at a time 
		and the output qubit is measured or lost. 
		The processing unit (PU) shifts the interferometric phase by $\Phi$ after each 
		successful measurement prior to processing the next qubit. 
	}
	\label{PAPER_ArXiv_2011:fig:MachZehnderInterferometer}
\end{figure}

We assume that the interferometric transformation (Fig.\ \ref{PAPER_ArXiv_2011:fig:MachZehnderInterferometer}),
can be expressed as a tensor product of quantum channels (i.e.\ completely\hyp{}positive trace\hyp{}nonincreasing maps \cite{Hou:J.Phys.A:2010}) $\mathcal{C}(\varphi;\Phi_m)$ for $\varphi$ the unknown phase shift being estimated and $\Phi_m$ a controllable phase with $m=0,1,\ldots,N-1$. 
The channel $\mathcal{C}$ is a noisy version of the restrictive single-qubit unitary process $U$
normally considered in QEM. Our tensor-product description corresponds to the 
assumption that the interferometric process, other than the control, is unchanging during the measurement procedure.
Photons of the $N$-qubit input state $\ket{\Psi_N}$ enter the interferometer one-by-one,
are transformed by $\mathcal{C}$.
Detectors measure where each photon exits, thereby implementing 
a projective\hyp{}valued measure with elements $\{\ket{0}\bra{0}, \ket{1}\bra{1}\}$ that 
yield one bit $u\in\{0,1\}$ if the photon is not lost.
The processing unit (PU) modifies the interferometric phase shift by $\Phi_m$,
according to the measurement history $h_m = u_m u_{m-1}\ldots u_1 \in \{0,1\}^m$ up to the $m^\text{th}$ photon,
prior to the next photon being processed.
After all $N$ input qubits have passed through the interferometer,
the PU estimates the interferometric phase shift $\varphi$ as $\widetilde{\varphi}$. 
A policy $\policyPRL$ is a `behavior pattern' for the PU, i.e., a collection of rules that 
tell the PU how to set $\Phi_m$ given $h_m$ and which phase estimate to report at the end.

The error probability distribution $P(\varsigma|\policyPRL)$ of the policy $\policyPRL$ yields the 
standard error $\Delta\varphi(\policyPRL)$ of the estimate $\widetilde{\varphi}$
for $\varsigma:= \varphi - \widetilde{\varphi}$.
As $\varsigma$ is cyclic over $2\pi$, $\Delta\varphi(\policyPRL)$ is given by the Holevo variance 
\makebox{$V_\text{H}(\policyPRL) = \Delta\varphi(\policyPRL)^2 :=S(\policyPRL)^{-2}-1$}, for 
$S(\policyPRL):=|\int_{-\pi}^\pi P(\varsigma|\policyPRL)\text{e}^{\text{i}\varsigma}\text{d}\varsigma|$ 
the sharpness of $P(\varsigma|\policyPRL)$ \cite{Holevo1984}. 
Evaluating $S(\policyPRL)$ requires exponential computing time with respect to $N$ and thus is computationally intractable. 
However, from $K$ trial runs of $\policyPRL$ with randomly chosen phases $\varphi_1,\ldots,\varphi_K$, we can infer a sharpness estimate 
$\tilde{S}:=|\sum_{k=1}^K\exp(\text{i}\varsigma_k)|/K$ for $\varsigma_k$ the error of the $k^\tn{th}$ phase estimate.
For QEM, $\Delta\varphi(\policyPRL)$ should scale better than the SQL $\Delta\varphi \propto 1/\sqrt{N}$
and as close as possible to the ultimate Heisenberg limit $\Delta\varphi \propto 1/N$ 
\cite{Luo:Lett.Math.Phys:2000,giovannetti:010401,Pezze&Smerzi:PhysRevLett.2009}.

For unitary evolution, the interferometer transforms each input qubit by
$U_{\bm{n}}(\theta)=\exp\{-\text{i} \theta\hat{\bm{\sigma}}\cdot\bm{n}\}$
for $\hat{\bm{\sigma}}:=(\hat{\sigma}_x, \hat{\sigma}_y, \hat{\sigma}_z)$ the Pauli matrices, 
$\bm{n}$ a unit vector, and $\varphi - \Phi = 2\theta$ the interferometric phase difference.
Without loss of generality, we can restrict our analysis to $\bm{n} = (0,1,0)$. 
However, because of imperfections, a real\hyp{}world interferometer 
is represented by a non\hyp{}unitary quantum channel $\mathcal{C}$.
We assume an unbiased interferometer, i.e.\
a random input qubit $\mathbbm{1} = \ket{0}\bra{0} + \ket{1}\bra{1}$ is mapped to itself
($\mathcal{C}(\mathbbm{1}) = \mathbbm{1}$), corresponding to a \emph{unital} channel. 
Hence, for continuous or discrete and \makebox{countable $\bm{n}$ and $\theta$ \cite{Mendl&Wolf:ComMathPhys2009},}
\begin{align}
	\mathcal{C}(\bullet)= \sumint\limits_{\bm{n},\theta}w_{\bm{n}}(\theta)U_{\bm{n}}(\theta)\bullet U_{\bm{n}}^\dagger(\theta), \quad w_{\bm{n}}(\theta) \in\mathbbm{R},
\end{align}
with $\sumint_{\bm{n},\theta}w_{\bm{n}}(\theta)=1$ and $w_{\bm{n}}(\theta) = \delta_{\theta,\varphi-\Phi}\delta_{\bm{n},(0,1,0)}$ for 
an ideal interferometer. 
In contrast, $\sumint_{\bm{n},\theta}w_{\bm{n}}(\theta)=1-\eta$ corresponds to an input state-independent loss rate $\eta$, and 
quantum noise is incorporated by $w_{\bm{n}}(\theta)$ being a general distribution with $\langle\theta\rangle=(\varphi-\Phi)/2$ and $\langle\bm{n}\rangle=(0,1,0)$. 
We simulate noise using  normal distributions with the aforementioned means
and small standard deviations \makebox{$\sigma_{\theta}, \sigma_{\bm{n}} \ll 1$,}
corresponding to visibility $1/(2 \text{e}^{2\sigma_{\theta}^2}-1)$.
For an optical interferometer,  $\theta$ noise corresponds to path-length difference fluctuations
and  $\bm{n}$ to beam splitter reflectivity fluctuations.
We utilize the input state 
\begin{equation*}
	\ket{\Psi_N}=\sum_{n,k= 0}^{N} \frac{\sin\hs{-2pt}\left(\frac{k+1}{N+2}\pi\right)}{\sqrt{1 + N/2}}\,
		\tn{e}^{\frac{\text{i}}{2}\pi(k-n)}d_{n-\frac{N}{2},k-\frac{N}{2}}^{N/2}\hs{-2pt}\left(\minfrac{\pi}{2}\right)\ket{n}_{[N]}
\end{equation*}
from \cite{PhysRevA.63.053804,PhysRevLett.85.5098,QLearning:hentschel:PRL:2010}, with $d_{\nu,\mu}^{j}(\beta)$ Wigner's $d$-matrix \cite{Group_theory_and_its_application_to_QM.Wigner.1931}. 
$\ket{n}_{[N]}$ is a permutationally-symmetric state with $n$ qubits in 
$\ket{1}$ and $N-n$ in $\ket{0}$ \cite{Hentschel:Permutationally-Symmetric_Qubit_Strings:J.Phys.A:2011}. 
The state $\ket{\Psi_N}$ is appealing 
because it allows precision close to the Heisenberg limit 
\cite{PhysRevA.63.053804,PhysRevLett.85.5098} and is robust against loss 
\cite{QLearning:hentschel:PRL:2010}, but our learning methods work for
other states as well.

\begin{figure}[t]
       \includegraphics[width=0.48\textwidth]{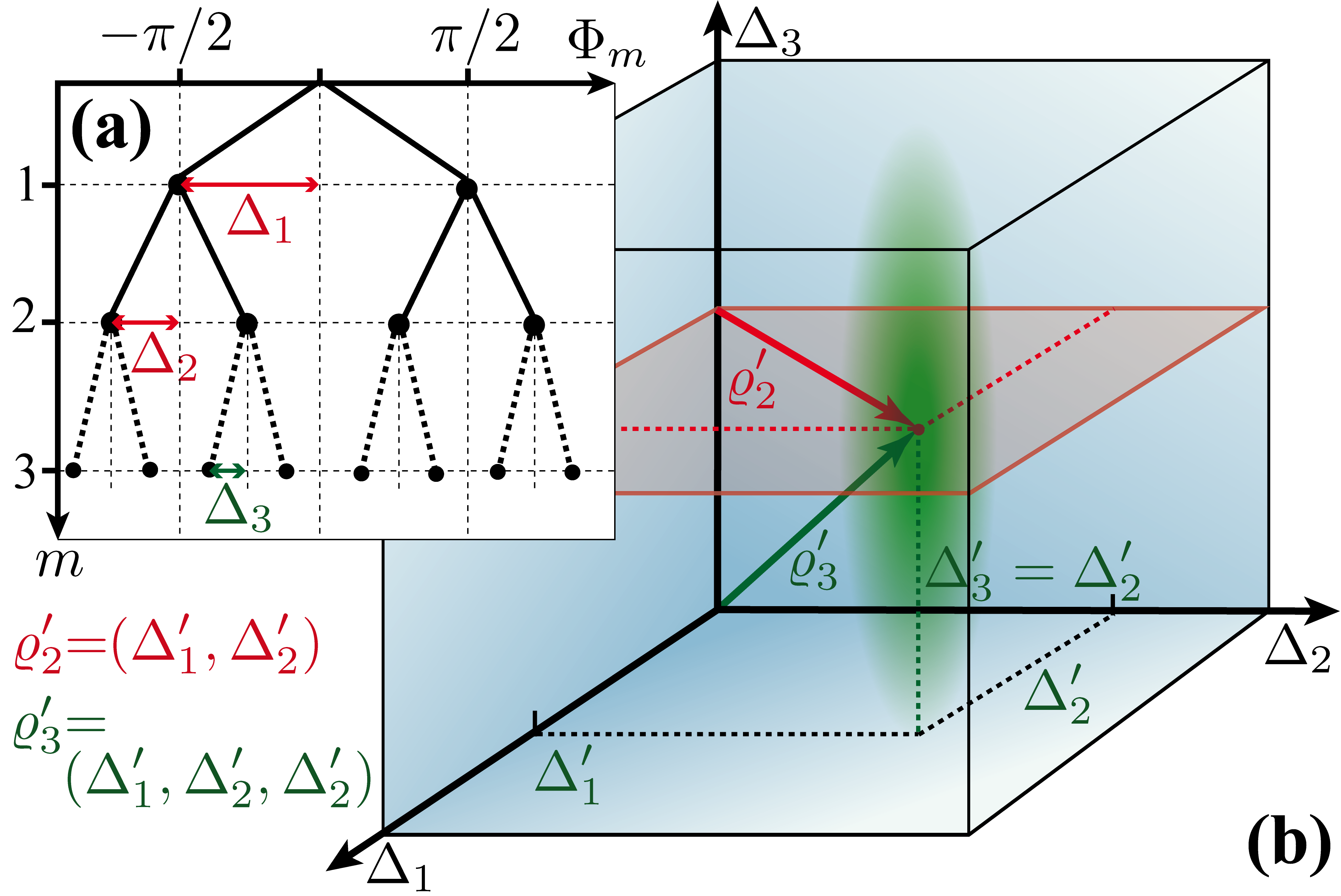}
	\caption[]{
		(a) Decision tree representation of a GLS policy for $N=2$ 
		(solid) and $N=3$ (entire tree). 
		For each path in the tree, the inner nodes represent the applied 
		feedback phases $\Phi_m$ and the leaf shows the final phase estimate $\pEst$.
		At depth $m$, a measurement $u_{m+1}=0$ directs the path to the left and 
		$u_{m+1}=1$ to the right. 
		(b) Embedding the best policy 
		$\policyPRL'_N\in\mathcal{P}_N$
		in the policy space $\mathcal{P}_{N+1}$,
		shown for $N=2$. From the best two-qubit policy 
		$\policyPRL'_2$, the policy 
		$\policyPRL'_3\in\mathcal{P}_3$ 
		is generated as a guideline. 
		The initial candidate policies for three input qubits are chosen according to 
		probability density \eqref{PAPER_ArXiv_2011:eq:PDF_for_policy_initialization}, 
		indicated by the shaded area around $\policyPRL'_3$.
 		(For clarity, the $N=2$ case is depicted,
		although only candidate policies for $N > 10$ are chosen according to 
		\eqref{PAPER_ArXiv_2011:eq:PDF_for_policy_initialization}.)
	}
	\label{PAPER_ArXiv_2011:fig:DecisionTree_and_Embedding}
\end{figure}

The control flow graph of any deterministic policy for a lossless conditions and a fixed $N$-qubit input state
can be represented as a binary decision tree of depth $N$ with an example shown in 
Fig.\ \hyperref[PAPER_ArXiv_2011:fig:DecisionTree_and_Embedding]{\ref{PAPER_ArXiv_2011:fig:DecisionTree_and_Embedding}(a)}.
Each of the $\sum_{\ell=0}^N 2^\ell = 2^{N+1}-1$ nodes of the tree corresponds to one specific state of 
the experiment and represents the resultant action of the policy.
Numeric optimization is computationally intractable due to the exponentially large number of nodes.
Therefore, we restrict our search to policies that implement a 
`generalized logarithmic search' (GLS) heuristic as described below, because
the set of all GLS policies can be parametrized by  only $N$ parameters and contains phase estimation
policies with optimal precision scaling \cite{QLearning:hentschel:PRL:2010} with respect to $N$.

For a uniform prior of $\varphi\in [0,2\pi)$,
the GLS heuristic commences with the initial feedback $\Phi_0=0$. 
After the $m^\tn{th}$ measurement result $u_m\in\{0,1\}$, the feedback phase is 
$\Phi_m = \Phi_{m-1} -(-1)^{u_m}\Delta_m$.
If the qubit is lost, $\Phi$ remains unchanged. 
After all $N$ input qubits are processed, there are $M\leq N$ measurement results 
$u_M,\ldots,u_1$, and the GLS heuristic reports the phase estimate
$\pEst = \Phi_{M-1} -(-1)^{u_M} \Delta_M$.
According to this parametrization, every GLS policy for an $N$-qubit input state is represented by a vector 
$\policyPRL =(\Delta_1,\ldots,\Delta_N)$ in the policy space $\mathcal{P}_N = [-\pi,\pi)^{N}$, and any such 
vector $\policyPRL \in\mathcal{P}_N$ is a valid policy. 
As any policy $\policyPRL \in\mathcal{P}_N$ utilizes a string of $N$ input qubits, we refer to it as an $N$-qubit policy.
Every $\policyPRL \in\mathcal{P}_N$ implements a GLS because $\policyPRL$ 
has variable entries compared to logarithmic search (LS) for which $\Delta_m = \frac{1}{2}\Delta_{m-1}$ \cite{Nowak:GLS:2008}.
The $N$-qubit LS policy $(\pi/2,\pi/4,\ldots,\pi/2^{N}) \in \mathcal{P}_N$ but does not surpass the SQL.
The duality between GLS policies and points in $\mathcal{P}_N \subset \mathbbm{R}^N$ allows
the use of function optimization techniques to search 
for an optimal $\policyPRL_\text{opt} \in \mathcal{P}_N$ with minimum $\Delta\varphi$, i.e.\ 
$\policyPRL_\text{opt}\in\operatorname*{arg\,min}_{\policyPRL\in\mathcal{P}_N} V(\policyPRL) 
  = \operatorname*{arg\,max}_{\policyPRL\in\mathcal{P}_N} S(\policyPRL)$.
Unfortunately, this optimization problem is non-convex and hence difficult \cite{QLearning:hentschel:PRL:2010}.

Particle swarm optimization (PSO) algorithms 
\cite{Eberhart1,Engelbrecht:2006:Computational_Swarm_Intelligence} are 
outstandingly successful for non-convex optimization. PSO is a `collective intelligence' strategy 
from the field of machine learning that learns via trial-and-error and performs as well as or 
better than simulated annealing and genetic algorithms
\cite{SA_vs_PSO_Ethni:2009,Kennedy98matchingalgorithms,Groenwold:2002}.
We have shown that PSO also 
delivers an autonomous approach to devising adaptive phase-estimation policies
for ideal interferometry \cite{QLearning:hentschel:PRL:2010,QLearning:hentschel:ITNG:2010}.

To search for $\policyPRL_\text{opt}$, the PSO algorithm models a `swarm' of $\swarmsize$
`particles' $\{p^{(1)},p^{(2)},\ldots,p^{(\swarmsize)}\}$ that move in the search space $\mathcal{P}_N$.
A particle's position $\policyPRL^{(i)} \in\mathcal{P}_N$ represents a candidate policy 
for estimating $\varphi$, which is initially chosen at random. 
Furthermore, $p^{(i)}$ remembers the best position, $\hat{\policyPRL}^{(i)}$, 
it has visited so far (including its current position). 
In addition, $p^{(i)}$ communicates with other particles in its neighborhood 
$\mathscr{N}^{(i)} \subseteq \{1,2,\ldots,\swarmsize\}$. 
We adopt the common approach to set each $\mathscr{N}^{(i)}$ in a pre-defined way regardless of the particles' positions
by arranging them in a ring topology: 
for $p^{(i)}$, all particles with maximum distance $r$ on the ring are in $\mathscr{N}^{(i)}$.
In iteration $t$, the PSO algorithm updates the position of all particles in a round-based manner as follows. 
\begin{enumerate}[label=(\roman*),widest=iii,  leftmargin=*,  topsep=0pt,partopsep=0pt,itemsep=0pt,parsep=0pt]
\item  Each particle $p^{(i)}$ samples $\tilde{S}(\policyPRL^{(i)})$ of its current position with $K$ trial runs. 
\item $p^{(i)}$ re-samples $\tilde{S}(\hat{\policyPRL}^{(i)})$ of its personal-best policy $\hat{\policyPRL}^{(i)}$,
	and the performance of $\hat{\policyPRL}^{(i)}$ is taken to be the arithmetic mean $\bar{S}(\hat{\policyPRL}^{(i)})$ 
	of all sharpness evaluations.
\item Each $p^{(i)}$ updates $\hat{\policyPRL}^{(i)}$ if $\tilde{S}(\policyPRL^{(i)}) > \bar{S}(\hat{\policyPRL}^{(i)})$ and
\item  communicates $\hat{\policyPRL}^{(i)}$ and $\bar{S}(\hat{\policyPRL}^{(i)})$ 
	to all members of $\mathscr{N}^{(i)}$.
\item Each particle $p^{(i)}$ determines the sharpest policy $\bm{\Lambda}^{(i)} = \max_{j \in \mathscr{N}^{(i)}} \hat{\policyPRL}^{(j)}$
	found so far by any one particle in $\mathscr{N}^{(i)}$ (including itself) and
\item  moves to
\begin{align}\label{PAPER_ArXiv_2011:eq:PSO-Base-Eq}
  \begin{split}
    \policyPRL^{(i)}	& \leftarrow \policyPRL^{(i)} +  \omega \bm{\delta}^{(i)}, \\
    \bm{\delta}^{(i)}	& \leftarrow \bm{\delta}^{(i)} + \beta_1\xi_1(\hat{\policyPRL}^{(i)} -\policyPRL^{(i)}) + \beta_2\xi_2(\bm{\Lambda}^{(i)} -\policyPRL^{(i)}).
  \end{split}
\end{align}
\end{enumerate}
The arrows indicate that the right value is assigned to the left variable.
The damping factor $\omega$ assists convergence,
and $\xi_1,\xi_2$ are uniformly-distributed random numbers from the interval $[0,1]$ that are re-generated each time Eq.\ \eqref{PAPER_ArXiv_2011:eq:PSO-Base-Eq} is evaluated.
The `exploitation weight' $\beta_1$ parametrizes the attraction of a particle to its personal best position $\hat{\policyPRL}^{(i)}$,
and the `exploration weight' $\beta_2$ describes attraction to the best position $\bm{\Lambda}^{(i)}$ in the neighborhood. 
To improve convergence, we bound each component of $\omega\bm{\delta}^{(i)}$ by a maximum value of $\nu_{\tn{max}}$. 
The user-specified parameters $\omega, \beta_1,\beta_2$, and $\nu_{\tn{max}}$ determine the swarm's behavior. 
Tests indicate that $\omega = 0.8$, $\beta_1 = 0.5$, $\beta_2 = 1$, 
and $\nu_{\tn{max}} = 0.2$ result in the highest probability to find an optimal policy.

\begin{figure}[!h]
\begin{minipage}[t]{0.48\linewidth}
   \centering
   \includegraphics[width=\linewidth]{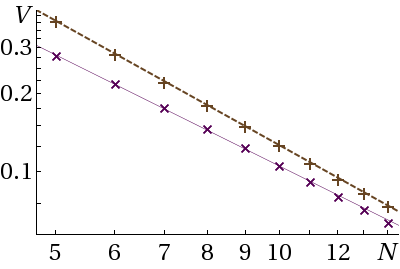}
   \vs{-18pt}
   \caption{ 
	Holevo phase variance $V_\text{H}$ of PSO optimized GLS policies. 
	Purple crosses (\textcolor{maroon}{\protect\scalebox{1.2}{$\times$}}): $N$-qubit policy used with input state $\ket{\Psi_N}$. 
	Brown pluses (\textcolor{brown}{\protect\scalebox{0.8}{$\bm{+}$}}): $(N-1)$-qubit policy used with input 
	state $\ket{\Psi_N}$ (the last measurement result is ignored by the policy). 
   }\label{PAPER_ArXiv_2011:fig:Performance_of_policy_for_N_to_N_plus_1}
\end{minipage}
\hspace{0.5cm}
\begin{minipage}[t]{0.48\linewidth}
   \centering
   \includegraphics[width=\linewidth]{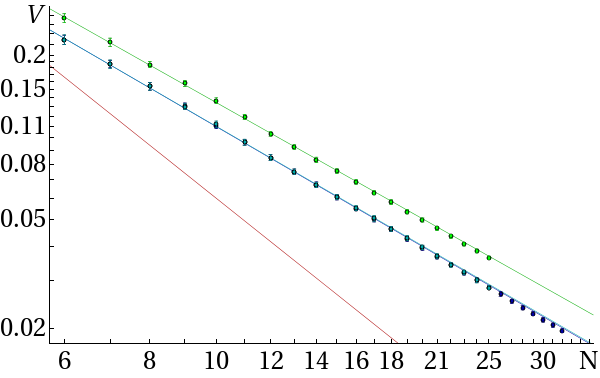}
   \vs{-18pt}
   \caption{
	  Holevo phase variance $V_\text{H}$ of policies optimized for 
	  simulated Gaussian quantum noise (\textcolor{green}{$\bullet$}) and 
	  skew-normal quantum noise with skewness $\gamma = 0.667$ (\textcolor{blue}{$\bullet$}).
	  In both cases, we used the standard deviations 
	  $\sigma_\theta = 0.02\pi$ and $\sigma_{\bm{n}_x} = \sigma_{\bm{n}_y} = \sigma_{\bm{n}_z} = 0.2\sigma_\theta$.
   }\label{PAPER_ArXiv_2011:fig:PSO-policies_for_Skew-Normal_Dist}
\end{minipage}
\end{figure}

The $K$ trial runs for assessing sharpness can be simulated or performed with a real world-experiment. 
For finite $K$, the sampled sharpness has statistical errors that can prevent the PSO algorithm from 
learning optimal solutions \cite{Bartz-Beielstein&el:Metaheuristics:2007}. 
We reduce sharpness errors by averaging over multiple samples in step (ii) \cite{SWIS-CONF-2005-004}. 
However, for $N > 12$, the PSO algorithm fails to learn good policies from scratch due to sharpness errors \cite{QLearning:hentschel:ITNG:2010}.
Therefore, we maintain our earlier strategy of running the learning algorithm for each $N$ independently when $N \leq 10$. 
For $N > 10$, our new heuristic bootstraps a starting point for the optimization of an $N$-qubit policy 
from the best $(N-1)$-qubit policy $\policyPRL' = (\Delta'_1,\ldots,\Delta'_{N-1})$.
Our heuristic exploits the fact that an $(N-1)$-qubit policy can be used as an $N$-qubit policy by ignoring the $N^\tn{th}$ measurement result. 
For $N \geq 10$, the optimal  $(N-1)$-qubit policy estimates phases with only $10\%$ less accuracy compared to an optimal $N$-qubit policy when used with the $N$-qubit input $\ket{\Psi_N}$
\footnote{
  See Fig.\ \ref{PAPER_ArXiv_2011:fig:Performance_of_policy_for_N_to_N_plus_1}, page \pageref{PAPER_ArXiv_2011:fig:Performance_of_policy_for_N_to_N_plus_1}.
}. 
Furthermore, the performance difference between the optimal $N$-qubit policy and the $(N-1)$-qubit policy decreases with increasing $N$ because the relative change in qubit number decreases with increasing $N$. 
Therefore, a good $(N-1)$-qubit policy is a valuable starting point for optimizing an $N$-qubit policy.


Utilizing previously learned policies is done at the initialization step of the PSO algorithm. 
The initial policy $\policyPRL \in \mathcal{P}_N$  
is selected as the particle's starting position with probability
\begin{align}\label{PAPER_ArXiv_2011:eq:PDF_for_policy_initialization}
  P(\policyPRL_{N}) =& 	
	\bigg( \prod_{k=1}^{N-1}\mathcal{N}_{\Delta'_k, \sigma_1}
	(\Delta_k)\bigg) \mathcal{N}_{\Delta'_N, \sigma_2}(\Delta_{N}),\\
  \mathcal{N}_{\mu, \sigma}(x) :=&
	\begin{cases}
		\kappa^{-1}_{\mu,\sigma}\exp\left\{ -\minfrac{(x-\mu)^2}{2\sigma^2}\right\},
			& x\in [0,\pi) 		\\
				0, 									& x\notin [0,\pi), 	\\
	\end{cases}\\
	\kappa^{-1}_{\mu,\sigma} 
		=&   \sqrt{\pi/2}\sigma\left[\operatorname{erf}\left(\frac{\pi-\mu}{\sqrt{2}\sigma}\right) 
			+ \operatorname{erf}\left(\frac{\mu}{\sqrt{2}\sigma}\right)\right],
\end{align}
with $\mathcal{N}_{\mu, \sigma}(x)$ a truncated normal distribution.
See Fig.\ \hyperref[PAPER_ArXiv_2011:fig:DecisionTree_and_Embedding]{\ref{PAPER_ArXiv_2011:fig:DecisionTree_and_Embedding}(b)} for
an illustration of this strategy.
The standard deviation $\sigma_1$ determines the similarity of the first $N$ actions of the newly 
generated policies compared to the template policy $\policyPRL'$. 
$\sigma_2$ determines the extent to which the action
for the new $N^\tn{th}$ qubit agrees with the previous 
action of $\policyPRL'$. 
We found that $\sigma_1 = 0.01\pi$ and $\sigma_2 = 0.25\pi$ yields a high success rate for our PSO heuristic. 

For $4 \leq N \leq 14$ and perfect interferometry, we verified that our new PSO algorithm with swarm size $\swarmsize = 20 N$ learns optimal $N$-qubit policies regardless of whether each policy's sharpness is evaluated exactly (requires time $\propto 2^N$) or sampled from $K = 10 N^2$ trial runs (requires polynomial runtime in $N$ when simulated).
Therefore, we sample the sharpness of each particle's current position and personal best position in each PSO iteration.
As we run the PSO algorithm for a constant $300$ iterations, 
the entire optimization process requires $\mathcal{O}(K\swarmsize)$ trials. 
However, to obtain an $N$-qubit policy, we have to optimize policies for $10,11,\ldots,N-1$ input qubits beforehand, 
as our algorithm requires an $(N-1)$ qubit policy for devising an $N$-qubit policy for any $N>10$. 
Therefore, learning an $N$-qubit policy requires $\mathcal{O}(N K\swarmsize) = \mathcal{O}(N^4)$ trial runs. 
When the trials are simulated, the computational complexity of our PSO heuristic is $\mathcal{O}(N^6)$
(hence efficient)
as a single trial run can be simulated in time $\mathcal{O}(N^2)$ \cite{Hentschel:Permutationally-Symmetric_Qubit_Strings:J.Phys.A:2011}.
Once learned, the execution of an $N$-qubit policy requires $N$ entangled input qubits.

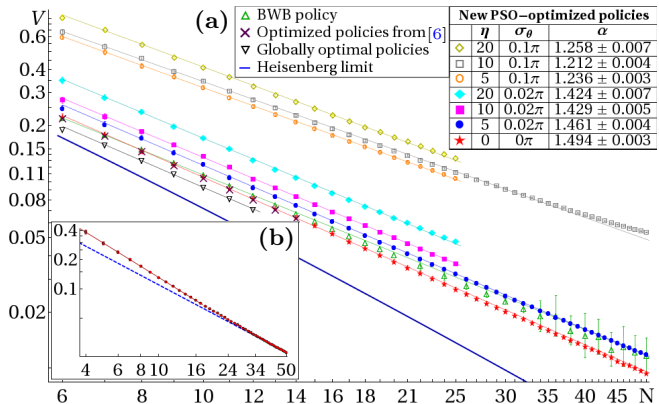
\begin{figure}
   \centering
   \begin{picture}(1,150)(122,5)
 		\put(0,0){\includegraphics[width=1\linewidth]{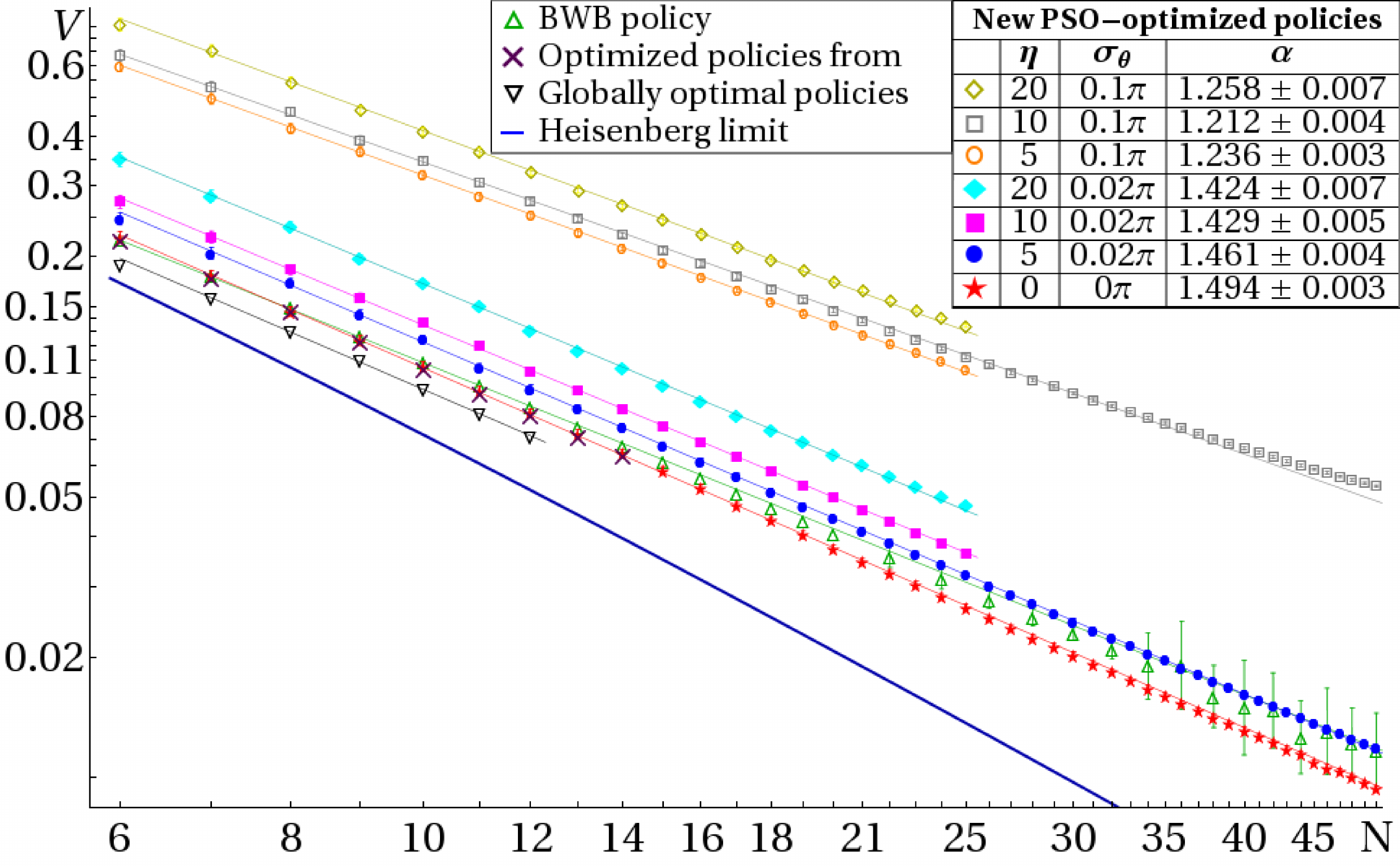}}
 		\put(71,144){\textbf{(a)}}
	 	\put(17,11){\includegraphics[width=0.38\linewidth]{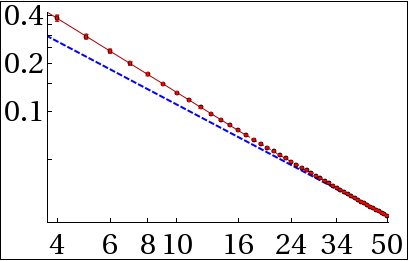}}
 		\put(95,62){\textbf{(b)}}
 		\put(159,140){\fontsize{6}{6}\selectfont\cite{QLearning:hentschel:PRL:2010}}
   \end{picture}
   \caption[A]{
	Holevo phase variance $V_\text{H}$ of PSO-optimized policies compared
	to other schemes vs.\ the number of input qubits $N$
	for (a) $\ket{\Psi_N}$ and (b) $\ket{0\cdots 00}$ as input states, respectively.
	The dashed line shows the SQL.
	Due to limited computational resources, some simulations are carried out only to $N\leq 25$. 
	(Loss rate $\eta$ is in percent; 
	$\sigma_{\bm{n}_x} = \sigma_{\bm{n}_y} = \sigma_{\bm{n}_z} = 0.2\sigma_\theta$.)
   }\label{PAPER_ArXiv_2011:fig:Final_Results}
\end{figure}

We trained our PSO algorithm with simulated trial runs for 
various noise and loss rates.
In each case, our PSO algorithm tries to find the sharpest policy 
$\policyPRL_N$ for given $N$.
As the algorithm uses stochastic optimization, it  is not guaranteed to learn the 
optimal policy every time and must be run several times independently for each $N$.
Nevertheless, within the limits of available computational resources, 
the PSO algorithm succeeded in at least 25\% of the runs, independently of $N$. 
We compared the policies generated by our new machine-learning algorithm to 
our previous numerically-optimized policies \cite{QLearning:hentschel:PRL:2010},
the Berry-Wiseman (BW) policy \cite{PhysRevLett.85.5098}, 
and policies obtained by  brute-force numerical optimization \cite{PhysRevA.63.053804}.

We first discuss policies for a noiseless, lossless setup, i.e., for unitary evolution.
Fig.\ \hyperref[PAPER_ArXiv_2011:fig:Final_Results]{\ref{PAPER_ArXiv_2011:fig:Final_Results}(a)} shows that our new method,
tested to the limits of available computational resources,
outperforms the BW-policy. 
We estimate the performance difference by calculating the scaling 
$\alpha$ of the Holevo variance $V_\text{H}$.
Our policies yield $V_\text{H}\propto N^{-\alpha}$ with $\alpha_\tn{PSO} = {1.494 \pm 0.003}$, 
compared to the inferred scaling $\alpha_\tn{BW} = {1.415 \pm 0.003}$ for $N \leq 50$.%
Furthermore, our new \emph{efficient} method greatly surpasses our previous optimization scheme \cite{QLearning:hentschel:PRL:2010} 
by more than tripling the domain of $N$ for developing policies while maintaining the same precision. 
The inefficient brute-force optimization was carried out in the full policy space, i.e.\ without restriction to GLS-policies. 
However, the resulting globally optimal policies perform better only by a constant factor of $0.88 \pm 0.01$ compared to our 
PSO-optimized policies but do not yield better scaling $\alpha$.
As expected the PSO algorithm yields policies approaching the SQL $V_\text{H} \propto 1/N$
for separable input states (Fig.\ \hyperref[PAPER_ArXiv_2011:fig:Final_Results]{\ref{PAPER_ArXiv_2011:fig:Final_Results}(b)})  \cite{Luo:Lett.Math.Phys:2000,giovannetti:010401,Pezze&Smerzi:PhysRevLett.2009}.

\begin{figure}[!h]
  \begin{minipage}[t]{0.48\linewidth}
    \begin{picture}(250,200)(0,-5)
	\put(0,0){\includegraphics[width=\linewidth]{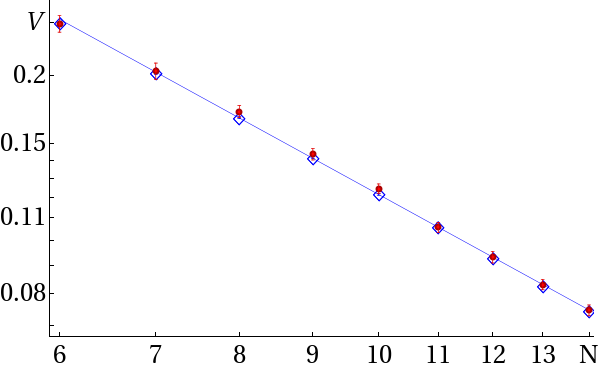}}
	\put(220,140){\bf \large (a)}
     \end{picture}%
  \end{minipage}
  \hspace{0.5cm}
  \begin{minipage}[t]{0.48\linewidth}
    \begin{picture}(250,200)(0,-5)
	\put(0,0){\includegraphics[width=\linewidth]{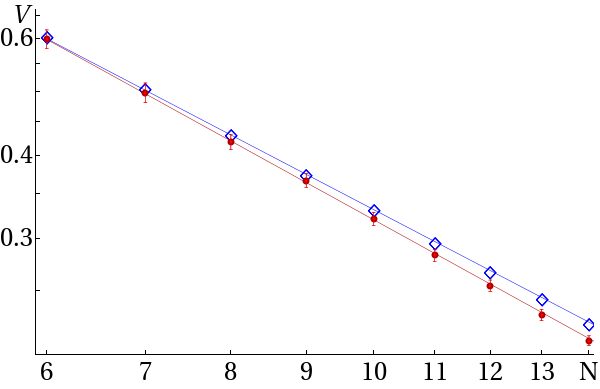}}
	\put(220,140){\bf \large (b)}
     \end{picture}%
  \end{minipage}
   \vs{-18pt}
   \caption{
	Holevo phase variance~$V_\text{H}$ of policies 
	from \cite{QLearning:hentschel:PRL:2010}, that are optimized for a perfect interferometer (\textcolor{blue}{$\Diamond$}),
	compared to the policies optimized by our new algorithm for the specific imperfections (\textcolor{darkred}{$\bullet$}). 
	The performance of the policies are evaluated for Gaussian quantum noise with standard deviations 
	  $\sigma_\theta$ and 
	  $\sigma_{\bm{n}} = (\varepsilon, \sqrt{1- \varepsilon^2},\varepsilon)$, $\varepsilon = 0.2\sigma_\theta$.
	(a) For low noise ($\eta = 5\%$ and $\sigma_\theta = 0.02\pi$), there is no noticeable performance enhancement.
	(b) For larger noise ($\eta = 5\%$ and $\sigma_\theta = 0.1\pi$), the policies optimized for perfect interferometry have a performance scaling $V_\text{H} \propto N^{-\alpha}$ with $\alpha_\tn{1} = {1.162 \pm 0.003}$. In contrast, the policies optimized for the aforementioned noise and loss achieve a scaling of $\alpha_\tn{2} = {1.236 \pm 0.003}$.
   }\label{PAPER_ArXiv_2011:fig:Performance_Difference_under_errors_and_loss}
\end{figure}

Our new algorithm delivers the first QEM policies optimized for a simulated imperfect interferometer with loss and Gaussian quantum noisy.
When applied to noisy conditions, policies generated by our new algorithm have significantly improved performances compared to policies optimized for perfect interferometry.
As expected, the performance difference increases with the noise level 
\footnote{
  See Fig.\ \ref{PAPER_ArXiv_2011:fig:Performance_Difference_under_errors_and_loss}, page \pageref{PAPER_ArXiv_2011:fig:Performance_Difference_under_errors_and_loss}.
}. 
We verify that our algorithm successfully devises superior policies also for non-Gaussian noise
by using skew-normal distributions with skewness $\gamma = 0.667$ 
for $P_\theta$ and $P_{\bm{n}}$ \cite{Azzalini:Scand.J.Stat.:2005}. 
We find that a nonzero third standardized moment with variances kept as before 
does not reduce the performance of the policies learned by our new PSO algorithm
\footnote{
  See Fig.\ \ref{PAPER_ArXiv_2011:fig:PSO-policies_for_Skew-Normal_Dist}, page \pageref{PAPER_ArXiv_2011:fig:PSO-policies_for_Skew-Normal_Dist}.
}.

In summary, we have devised an \emph{efficient} machine learning algorithm to construct adaptive-feedback 
measurement policies autonomously for time-independent, single-parameter estimation problems. 
Our one prerequisite is a training-phase comparison criterion to evaluate the success of 
candidate policies.
Within the limits of available computational 
resources, our PSO-generated policies outperform all known schemes for adaptive single-shot 
phase estimation with direct measurement of the channel output.
Our algorithm learns to account for experimental errors and loss
thereby making time-consuming error modeling and extensive calibration dispensable.
\vfill

\noindent
\emph{Acknowledgments:}
We thank B.\,Bunk and Humboldt-Universit\"at zu Berlin for computational resources, 
and D.\,W.\.Berry, L.\,Maccone, and H.\,M.\,Wiseman for comments on an earlier draft.
This project has been supported by \textit{i}CORE, AITF, NSERC, and CIFAR.
BCS is supported by a CIFAR Fellowship.

\onecolumngrid \newpage \twocolumngrid

\onecolumngrid \clearpage 

\begin{center}
	 {\bf \large Appendix}
\end{center}

\vs{20pt}

\noindent

\end{document}